\def\be{\begin{equation}}
\def\ee{\end{equation}}
\def\ba{\begin{array}}
\def\ea{\end{array}}

\documentclass[priprint,pra,showpacs,amsmath]{revtex4}
\usepackage{amssymb}
\usepackage{CJK}
\def\qed{\leavevmode\unskip\penalty9999 \hbox{}\nobreak\hfill
     \quad\hbox{\leavevmode  \hbox to.77778em{%
               \hfil\vrule   \vbox to.675em%
               {\hrule width.6em\vfil\hrule}\vrule\hfil}}
     \par\vskip3pt}

\input amssym.def
\begin{document}
\begin{center}\bf A Note on Local Unitary Equivalence of Isotropic-like States \end{center}
\vskip 1mm

\begin{center}
{Tinggui Zhang$^{1}$},
{Bobo Hua$^{2}$},
{Ming Li$^{3}$},
{Ming-Jing Zhao$^{4}$},
and {Hong Yang$^{5}$}

\vspace{2ex}

\begin{minipage}{5in}

\small $~^{1}$ {School of Mathematics and Statistics, Hainan Normal University, Haikou 571158, China}

{\small $~^{2}$ School of Mathematical Sciences, LMNS, Fudan University,
Shanghai 200433, China}

{\small $~^{3}$ College of the Science, China
University of
Petroleum, 266580 Qingdao, China}

{\small $~^{4}$ Department of Mathematics, School of Science, Beijing
Information Science and Technology University, 100192, Beijing,
China}

{\small $~^{5}$ College of Physics and Electronic Engineering, Hainan Normal
University, Haikou 571158, China}

\end{minipage}
\end{center}

{\bf Abstract} We consider the local unitary equivalence of a class
of quantum states in bipartite case and multipartite case. The
necessary and sufficient condition is presented.
 As special cases, the local unitary equivalent classes of isotropic state and werner state are provided.
 Then we study the local unitary similar equivalence of this class of quantum states and analyze the necessary and sufficient
 condition.

Pacs numbers: {03.67.-a, 02.20.Hj, 03.65.-w}

\bigskip


{ E-mail address}: zhaomingjingde@126.com (Ming-Jing Zhao)

\maketitle

\section{Introduction}

Entanglement is one of the most extraordinary features of quantum
physics. It plays a vital role in quantum information processing,
including quantum teleportation, quantum cryptography, quantum
computation, etc. \cite{EofCon}. One fact is that two entangled
states are said to be equivalent in implementing the same quantum
information task if they can be obtained from each other via local
operation and classical communication(LOCC). In particular, all the
LOCC equivalent quantum pure states are interconvertible by local
unitary operators (LU)\cite{Wdur}. As many properties like quantum
correlation, quantum entanglement, quantum discord keep invariant
under local unitary transformations, it is significant to classify and
characterize quantum states in terms of local unitary
transformations.

There are a lot of literatures to deal with the LU problem, one
approach is to construct invariants of local unitary transformations
\cite{Grassl,makhlin,linden, Linden,SFG, SFW, SFY,SCFW}. Usually the
invariants of mixed states are dependent of pure state
decomposition. Recently, the invariants of bipartite states
independent of the pure states decomposition are studied in \cite{tnxm}.
the LU problem for multipartite pure qubits
states has been solved in \cite{mqubit}. By exploiting the high
order singular value decomposition technique and local symmetries of
the states, Ref. \cite{bliu} presents a practical scheme of
classification under local unitary transformations for general
multipartite pure states with arbitrary dimensions, which extends
results of n-qubit pure states \cite{mqubit} to that of n-qudit pure
states. For mixed states, Ref. \cite{zhou} solved the LU problem of arbitrary dimensional bipartite
non-degenerated quantum systems by presenting a complete set of
invariants, such that two density matrices are locally unitary
equivalent if and only if all these invariants have equal values. In
\cite{zhang} the case of multipartite systems is studied and a
complete set of invariants is presented for a special class of mixed
states. Recently, we have studied the local unitary equivalence of
multipartite mixed states using the technology of matrix realignment
and partial transpose \cite{zhang1} and solved the LU problem for multi-qubit mixed states with Bloch
representation\cite{liwang}.

In this paper, we study the LU problem for a special class of quantum states. The necessary and sufficient condition is provided. Especially, the local unitary equivalence class of
isotropic states \cite{mphoro} and Werner states \cite{werner} are obtained. Then we study the local unitary similar equivalence of this class of states and give the necessary and sufficient condition.

\section{Local unitary equivalence}

Two multipartite mixed states $\rho$ and $\rho^\prime$ in
$H_1\otimes H_2\otimes \cdots \otimes H_n$ are said to be equivalent
under local unitary transformations if there exist unitary operators
$U_i$ on the $i$-th Hilbert space $H_i$ such that \be\label{eq}
\rho^\prime=(U_1\otimes U_2\otimes\cdots \otimes U_n)\rho(U_1\otimes
U_2\otimes \cdots \otimes U_n)^\dag. \ee

First, we consider the case of bipartite system. Let $H$ be an
$N-$dimensional complex Hilbert space with $|i\rangle$,
$i=1,2,\cdots,N$ an orthonormal basis. A general pure state on
$H\otimes H$ is of the form
\be\label{li01}|\phi\rangle=\sum_{i,j=1}^Na_{ij}|i\rangle\otimes|j\rangle,
\ \ a_{ij}\in \mathbb{C}\ee with the normalization
$\sum_{i,j=1}^Na_{ij}a_{ij}^\ast=1$($x^\ast$ denotes the complex
conjugation of $x$). Let $A$ denote the matrix given by
$(A)_{ij}=a_{ij}$, we call $A$ the matrix representation of
pure state $|\phi\rangle$. The following quantities are associated
with the state $|\phi\rangle$ given by
(\ref{li01}).\be\label{li02}I_\alpha=Tr(AA^\dag)^\alpha, \ \
\alpha=1,2,\cdots,N,\ee where $A^\dag$ denotes the adjoint of the
matrix $A$. It is well-known that two bipartite pure states
$|\phi_1\rangle$ and $|\phi_2\rangle$ in $H\otimes H$ are local
unitary equivalent if and only if their matrix representation give
the same values of quantities (\ref{li02}).

Here we mainly consider the local unitary equivalence of quantum states
\begin{eqnarray}\label{eq bi state-1}
\rho_1=\frac{p_0}{N^2}I_N\otimes I_N +\sum_{i=1}^K p_i
|\phi_i\rangle\langle\phi_i|
\end{eqnarray}
and
\begin{eqnarray}\label{eq bi state-2}
\rho_2=\frac{p_0}{N^2}I_N\otimes
I_N + \sum_{i=1}^K p_i |\varphi_i\rangle\langle\varphi_i|
\end{eqnarray}
with
$p_i\geq 0$ for $i=0, \cdots, K$, $\sum_{i=0}^K p_i=1$, and $p_i\neq
p_j$ for $1\leq i < j \leq K$, $1\leq K\leq N^2$.

\noindent {\bf Lemma 1:}
Two arbitrary dimensional bipartite non-degenerate density
matrices are equivalent under local unitary transformations
if and only if there exist eigenstate decompositions
$\rho=\sum_i p_i |\psi_i\rangle\langle\psi_i|$ such that the following invariants have the same values
for both density matrices:
\begin{eqnarray}
&&J^s=Tr_2(Tr_1 \rho^s), \ \ s=1,\cdots, N^2,\\
&&Tr[(A_iA_j^\dagger)(A_kA_l^\dagger)\cdots(A_hA_p^\dagger)].\label{eq lu some pure}
\end{eqnarray}

\noindent {\bf Proposition 1:} For two bipartite mixed states in Eq. (\ref{eq bi state-1}) and Eq. (\ref{eq bi state-2}), they are local
unitary equivalent if and only if the corresponding matrix
representations of $|\phi_i\rangle$ and $|\varphi_i\rangle$ yield
the same values of the invariants (\ref{eq lu some pure}).

\noindent {\bf Proof:}
If $\rho_1$ and $\rho_2$ are local unitary
equivalent, then $|\phi_i\rangle$ and $|\varphi_i\rangle$ are local
unitary equivalent under the same local unitary operators. Therefore, $|\phi_i\rangle$ and $|\varphi_i\rangle$
give rise to the same values of the invariants (\ref{eq lu some pure}).

On the other hand, if $|\phi_i\rangle$ and $|\varphi_i\rangle$ give
rise to the same values of the invariants (\ref{eq lu some pure}).
By Lemma 1, $|\phi_i\rangle$ and $|\varphi_i\rangle$ are local
unitary equivalent under the same local unitary operators, hence
$\rho_1$ and $\rho_2$ are local unitary equivalent.

{\bf Remark:} In fact, if the eigenvalues are not all positive in Proposition 1, then the conclusion still holds true. The Proposition 1 can be used to solve the local
unitary equivalence of mixed state with only one degenerate
eigenvalue. Because if one state has only one degenerate
eigenvalues, then it can be transformed to the form like Proposition
1. That is $\rho=\lambda_1|v_1\rangle\langle v_1|+ \cdots
+\lambda_s|v_s\rangle\langle
v_s|+\sum_{i=s+1}^{N^2}\lambda_0|v_i\rangle\langle v_i|$,
equivalently,
$\rho=\lambda_0I_{N^2}+(\lambda_1-\lambda_0)|v_1\rangle\langle v_1|+
\cdots +(\lambda_s-\lambda_0)|v_s\rangle\langle v_s|$, where
$\lambda_i\neq\lambda_j,\ \ i\neq j, \ i, j=0,1,\cdots,s$.

Now we can analyze the LU problem in two-qubit
system. First, when quantum state has non-degenerate eigenvalues,
then Lemma 1 is sufficient to determine the local unitary equivalence.
Second, when quantum state has eigenvalues with multiplicity not larger
than 2, then one can solve the local unitary equivalence by the
method proposed in \cite{zhang1}. At last, if there is only one
degenerate eigenvalue, then Proposition 1 can be used to deal with the
LU problem of quantum states. Therefore, the LU problem of two-qubit quantum states can be
solved in this way.

This Proposition can also be used to judge which states are
equivalent to isotropic states \cite{mphoro} under local unitary transformations, which are
invariant under transformations of the form $(U\otimes
U^\ast)\rho(U\otimes U^\ast)^\dag$. Isotropic state can be written as the
mixture of the maximally mixed state and the maximally entangled
state $|\psi^+\rangle=\frac{1}{\sqrt{d}}\sum_{a=0}^{d-1}|aa\rangle,$
$$\rho_{isot}=\frac{p}{d^2}I_d\otimes I_d + (1-p)|\psi^+\rangle\langle\psi^+|,$$
$0\leq p\leq 1$.
Following Proposition 1, the state that is local unitary equivalent
to the isotropic states is the form $\rho=\frac{p}{d^2}I_d\otimes
I_d + (1-p)|\psi^{\prime}\rangle\langle\psi^{\prime}|$, where
$|\psi^{\prime}\rangle$ is a maximally entangled state.

Subsequently, we consider the states that are local unitary equivalent to
Werner states \cite{werner}. We need the technique of partial
transpose of states. For a density matrix $\rho$ in $H_1\otimes H_2$
with elements $\rho_{m\mu,n\nu}=\langle e_m\otimes
f_\mu|\rho|e_n\otimes f_\nu\rangle$, the partial transposition of
$\rho$ is defined by \cite{mprh}:
$$
\rho^{T_2}=(I\otimes
T)\rho=\sum_{mn,\mu\nu}\rho_{m\nu,n\mu}|e_m\otimes
f_{\mu}\rangle\langle e_n\otimes f_{\nu}|,
$$
where $\rho^{T_2}$ denotes the transposition of $\rho$ with respect
to the second system, $|e_n\rangle$ and $|f_\nu\rangle$ are the
bases associated with spaces $H_1$ and $H_2$ respectively. The
LU problem of the original states can be transformed
to that of theirs partial transposed
states \cite{zhang1}, since two mixed states $\rho_1$ and $\rho_2$ in
$H_1\otimes H_2$ are local unitary equivalent if and only if
$\rho_1^{T_2}$ and $\rho_2^{T_2}$ are local unitary equivalent.

The arbitrary dimensional Werner states \cite{werner} are
invariant under the transformations $(U\otimes U)\rho(U\otimes
U)^\dag$ for any unitary $U$. They can be written as
$$\rho_w=\frac{1}{d^3-d}[(d-f)I_d\otimes
I_d+(df-1)\sum_{ij}|ij\rangle\langle ji|],$$
where $-1 \leq f \leq 1$. The partial transpose of
$\rho_w$ is
$\rho_w^{T_2}=\frac{1}{d^3-d}(d-f)I_d\otimes
I_d+\frac{(df-1)}{d^2-1}|\psi^+\rangle\langle \psi^+|$. Therefore, the state that is local unitary equivalent
to the werner states is of the form $\rho=\frac{1}{d^3-d}(d-f) I_d\otimes
I_d + \frac{(df-1)}{d^2-1}(|\psi^{\prime}\rangle\langle\psi^{\prime}|)^{T_2}$, where
$|\psi^{\prime}\rangle$ is a maximally entangled state.

Now we consider the multipartite case.
Before showing the equivalence of multipartite quantum states under
local unitary transformations, we give a short review of high order
singular value decomposition developed in \cite{ldbdj}. For any
tensor $\mathcal {A}$ with order $d_1\times d_2\times\cdots\times
d_N$, there exists a core tensor $\Sigma$ such that
\be \mathcal
{A}=(U_1\otimes U_2\otimes\cdots\otimes U_N)\Sigma,
\ee
where $\Sigma$ forms the same order tensor with $\mathcal {A}$. Any
$N-1$ order tensor $\sum_{i_n=i}$ obtained by fixing the n-th index
to $i$, has the following properties
$\langle\sum_{i_n=i},\sum_{i_n=j}\rangle=\delta_{ij}\sigma_i^{(n)^2},$
with $\sigma_i^{n}\geq \sigma_j^{n}$ and $ \forall i \leq j$ for all
possible values of $n$. Here, the singular value $\sigma_i^{n}$
symbolizes the Frobenius norm
$\sigma_i^{n}=\|\sum_{i_n=i}\|\equiv\sqrt{\langle\sum_{i_n=i},\sum_{i_n=i}\rangle}$,
where the inner product $\langle A,B
\rangle\equiv\sum_{i_1}\sum_{i_2}\cdots\sum_{i_N}b_{i_1i_2\cdots
i_N}a_{i_1i_2\cdots i_N}^\ast$

To calculate the core tensor $\Sigma$, one first expresses $\mathcal
{A}$ in matrix unfolding form $\mathcal {A}_n$. Then one derives the
singular value decomposition of the matrix $\mathcal
{A}_n=U_n\Lambda_n V_n$. The core tensor is then given
by \be \Sigma=\otimes_{n=1}^{N}U_n^{\dag}{\mathcal {A}}. \ee

\noindent {\bf Lemma 2:} Two multipartite pure states are local
unitary equivalent if and only if they have the same core tensor up
to the local symmetry $\otimes_{n=1}^N P^{(n)}$, where $P^{(n)}$ is a
block-diagonal matrix consisting of unitary blocks with the same
partitions as that of the identical singular values of $\mathcal {A}_n$.

By Proposition 1 and Lemma 2, one can get the following result easily.

\noindent {\bf Proposition 2:} Two multipartite mixed states of the
form $\rho_1=\frac{p}{NM\cdots T}I_N\otimes I_M \otimes\cdots\otimes
I_T+ (1-p)|\phi\rangle\langle\phi|$ and $\rho_2=\frac{p}{NM\cdots
T}I_N\otimes I_M \otimes\cdots\otimes I_T +
(1-p)|\varphi\rangle\langle\varphi|$ are local unitary equivalent if
and only if $|\phi\rangle$ and $|\varphi\rangle$ have the same core
tensor up to the local symmetry $\otimes_{n=1}^N P^{(n)}$.

\noindent {\bf Remark:} Two multipartite mixed states
$\rho_1=\frac{p}{NM\cdots T}I_N\otimes I_M \otimes\cdots\otimes
I_T+ (1-p)\rho$ and $\rho_2=\frac{p}{NM\cdots
T}I_N\otimes I_M \otimes\cdots\otimes I_T +
(1-p)\rho^\prime$ are local
unitary equivalent if and only if the corresponding density matrices
$\rho$ and $\rho^\prime$ are local unitary equivalent. Therefore, the local unitary equivalence of two quantum states does not change under the disturbance with the white noise. For example, $\rho_1=\frac{p}{8}I_2\otimes I_2\otimes
I_2+ (1-p)\rho$ and $\rho_2=\frac{p}{8}I_2\otimes I_2\otimes
I_2+ (1-p)\rho^\prime$ with $\rho=\frac{q}{2}(|000\rangle+|111\rangle)(\langle000|+\langle111|)+(1-q)|111\rangle\langle111|$ and $\rho^\prime=\frac{q}{2}(|001\rangle+|010\rangle+|100\rangle)(\langle001|+\langle010|+\langle100|)+(1-q)|111\rangle\langle111|$, are not local unitary equivalent because $\rho$ and $\rho^\prime$ are not local unitary equivalent \cite{zhang}.

\section{Local unitary similar equivalence}

\noindent {\bf Definition:} If there exists a unitary matrix $U$
such that $(U\otimes U^\ast)\rho_1 (U\otimes U^\ast)^\dag=\rho_2$,
we call states $\rho_1$ and $\rho_2$ local unitary similar
equivalent.

In \cite{hshapiro} the author studies the unitary invariants and
unitary similar equivalence, and the following Specht's theorem
\cite{wspe} has been presented. Next we use them to deal with the
local unitary similar equivalent problem for bipartite mixed states.

\noindent {\bf Lemma 3:} Let $A$ and $B$ be $n\times n$ complex
matrices. Then $A$ and $B$ are unitary similar, i.e  there is a
unitary matrix $U$, such that $UAU^\dag = B$, if and only if
$tr(\omega(A,A^\dag))=tr(\omega(B,B^\dag))$ holds for every word
$\omega$, where $\omega(A,A^{\dag})$ is the result of taking any
monomial $\omega(x,y)$ in noncommuting variables $x$ and $y$ and
replacing $x$ with $A$ and $y$ with $A^\dag$.

The proof of Specht's theorem can also be applied to two finite
sets $\{A_i\}_{i=1}^t$ and $\{B_i\}_{i=1}^t$ of $n\times n$ matrices
\cite{hshapiro,nwie}.

 \noindent {\bf Lemma
4:} Let $\{A_i\}_{i=1}^t$ and $\{B_i\}_{i=1}^t$ be $n\times n$
complex matrices. There is a unitary matrix $U$ such that $U^\dag
A_iU=B_i$ for $i=1,2,\cdots, t$ if and only if for every word
$\omega(x_1,y_1,x_2,y_2,\cdots,x_t,y_t)$ in the noncommuting
variables $x_i$ and $y_i$ we have
$tr(\omega(A_1,A_1^{\dag},A_2,A_2^{\dag},\cdots,A_t,A_t^{\dag}))=tr(\omega(B_1,B_1^{\dag},B_2,B_2^{\dag},\cdots,B_t,B_t^{\dag}))$.

For pure state $|\psi\rangle$ and $\phi\rangle$ with coefficient matrices $A$ and $B$ respectively, if $U\otimes U^*|\psi\rangle=|\phi\rangle$, then $UAU^{\dagger}=B$. Utilizing this relation, we can get the necessary and sufficient condition for local unitary similar equivalence problem .

\noindent {\bf Proposition 3:} For two bipartite mixed states in Eq. (\ref{eq bi state-1}) and Eq. (\ref{eq bi state-2}),
they are local unitary similar equivalent if and only if $tr(\omega(A_1,A_1^{\dag},A_2,A_2^{\dag},\cdots,A_K,A_K^{\dag}))=tr(\omega(B_1,B_1^{\dag},B_2,B_2^{\dag},\cdots,B_K,B_K^{\dag}))$ holds true for every word
$\omega(x_1,y_1,x_2,y_2,\cdots,x_t,y_t)$ in the noncommuting
variables $x_i$ and $y_i$, with $A_i$ and $B_i$ the coefficient matrices of $|\phi_i\rangle$ and $|\varphi_i\rangle$ respectively.

\section{Conclusions}

In summary, we have studied the LU problem
for a special class of states. The necessary and sufficient condition is provided. Consequently, the local unitary equivalent classes of isotropic state and werner state are obtained. Then we have investigated the local unitary similar equivalence for this class of state and obtained the necessary and sufficient condition.

\section{Acknowledgement}
We would like to thank Prof. Shao-Ming Fei and Xianqing Li-Jost for
useful discussions. This work is supported by the NSF of China under
Grant No. 11401032 and No. 61473325; the NSF of Hainan Province
under Grant No.20151005 and No.20151010; Scientific Research
Foundation for the Returned Overseas Chinese Scholars, State
Education Ministry.


\begin{thebibliography}{99}


\bibitem{EofCon} R. Horodecki, P. Horodecki, M. Horodecki, K. Horodecki,  Rev. Mod. Phys {\bf 81} (2009) 865.
\bibitem{Wdur}  W. ~D\"ur, G. Vidal, J.I. Cirac, Phys. Rev. A {\bf
62} (2000) 062314.

\bibitem{Grassl} M. ~Grassl,  M. R\"otteler,  T. Beth, Phys. Rev. A {\bf 58} (1998) 1833.

\bibitem{makhlin} Y. Makhlin, Quant. Info. Proc. {\bf 1} (2002) 243.

\bibitem{linden}  N. Linden, S. Popescu, A. Sudbery, Phys. Rev. Lett {\bf 83} (1999) 243.

\bibitem{Linden}  N. Linden, S. Popescu,  Phys {\bf 46} (1998) 567.

\bibitem{SFY}  S. Albeverio, S.M. Fei, P. Parashar,  W.L. Yang, Phys. Rev. A {\bf 68} (2003) 010303.

\bibitem{SFG}  S. Albeverio,  S.M. Fei, D. Goswami, Phys. Lett. A {\bf 340} (2005) 37.

\bibitem{SFW}  B.Z. Sun,  S.M. Fei, X. Li-Jost, Z.X. Wang, J. Phys. A {\bf 39} (2006) L43.

\bibitem{SCFW} S. Albeverio,  L. Cattaneo, S.M. Fei, X.H. Wang, Int. J. Quant. Inform {\bf 3} (2005) 603.
\bibitem {tnxm}  T.G. Zhang,  N. Jing,  X. Li-Jost, M.J. Zhao, S.M. Fei, Euro. Phys. J. D {\bf 67} (2013) 175.
\bibitem{mqubit}  B. Kraus, Phys. Rev. Lett. {\bf 104} (2010) 020504; Phys. Rev. A {\bf 82} (2010) 032121.

\bibitem{bliu} B. Liu, J.L. Li, X. Li, C.F. Qiao, Phys. Rev. Lett {\bf 108} (2012) 050501.

\bibitem{zhou}  C. Zhou,  T. Zhang, S.M. Fei,  N. Jing,  X. Li-Jost, Phys. Rev. A {\bf 86} (2012) 010303.

\bibitem{zhang}  T.G. Zhang,  M.J. Zhao,  X. Li-Jost, S.M. Fei, Int. J. Theor. Phys
{\bf 52} (2013) 3020.
\bibitem{zhang1}  T.G. Zhang, M.J. Zhao, M. Li, S.M. Fei, X. Li-Jost, Phys. Rev. A {\bf 88} (2013) 042304.
\bibitem{liwang}  M. Li, T.G. Zhang, S.M. Fei, X. Li-Jost, N. Jing, Phys. Rev. A {\bf 89} (2014) 062325.


\bibitem{mphoro}  M. Horodecki, P. Horodecki, Phys. Rev. A {\bf 59} (1999) 4206.
\bibitem{werner}  R.F. Werner, Phys. Rev. A {\bf 40} (1989) 4277.
\bibitem{mprh}  M. Horodecki, P. Horodecki, R.Horodecki, Phys. Lett. A {\bf 233} (1996) 1.

\bibitem{ldbdj}  L.D. Lathaumer, B.D. Moor, J. Vandewalle, SIAM. J. Matrix Anal. Appl {\bf 21} (2000) 1253.
\bibitem{hshapiro}  H. Shapiro, Linear Algebra and Appl. {\bf 147}(1991) 101.

\bibitem{wspe} W. Specht, Deutsch. Math.-Verein {\bf 50} (1940) 19.

\bibitem{nwie} N. Wiegmann, J. Austral. Math. Soc {\bf 2} (1962) 122 .


\end{thebibliography}
\end{document}